\documentclass[prl,twocolumn,amsmath,amssymb,superscriptaddress,showpacs]{revtex4-1}

\usepackage{graphicx}
\usepackage{epsfig}
\usepackage{epstopdf}

\begin{document}

\title{Controlling electron-phonon interactions in graphene at ultra high carrier densities}
\author{Dmitri K. Efetov}
\author{Philip Kim}
\affiliation{Department of Physics, Columbia University New York, NY 10027}
\date{\today}

\begin{abstract}
We report on the temperature dependent electron transport in graphene at
different carrier densities $n$. Employing an electrolytic gate,
we demonstrate that $n$ can be adjusted up to
4$\times$10$^{14}$cm$^{-2}$ for both electrons and holes. The
measured sample resistivity $\rho$ increases linearly with
temperature $T$ in the high temperature limit, indicating that a
quasi-classical phonon distribution is responsible for the electron
scattering. As $T$ decreases, the resistivity decreases
more rapidly following $\rho (T) \sim T^{4}$. This low temperature
behavior can be described by a Bloch-Gr\"{u}neisen model taking
into account the quantum distribution of the 2-dimensional acoustic phonons in graphene. We map out the density dependence of the characteristic temperature $\Theta_{BG}$ defining the cross-over between the two distinct regimes, and show, that for all $n$, $\rho(T)$ scales as a universal function of the normalized temperature $T/\Theta_{BG}$.
\end{abstract}

\pacs{73.63.b, 73.22.f, 73.23.b}
\maketitle

At finite temperatures electrons in typical conductors are scattered by
phonons, producing a finite, temperature dependent resistivity $\rho$~\cite{Abrikosov}. If the temperature $T$ is
comparable to or larger than the Debye temperature $\Theta_D$ --- the representative temperature scale for the highest phonon energies --- all
phonon modes are populated. In this high temperature regime, $\rho(T)\sim T$,
reflecting a classical equipartition distribution of the phonons. As $T$ decreases below
$\Theta_D$, however, the bosonic nature of the phonons becomes important : only the acoustic phonon modes within the phonon sphere of diameter $k_{ph}=k_BT/\hbar v_s<k_{D}$ (where $k_{D}$ is the radius of the Debye sphere and $v_{s}$ is the sound velocity) are populated appreciably, leading to a more rapid decrease of the resistivity, $\rho(T)\sim T^{5}$, known as the Bloch-Gr\"{u}neisen (BG) regime~\cite{Bloch,Grueneisen,Meisner1935}.

Due to the quasi-elasticity of the electron-phonon (e-ph)
interactions, the maximal phonon momentum in an e-ph scattering
event is limited to $2\hbar k_{F}$, representing a full
backscattering of the electrons across the Fermi surface of radius
$k_{F}$. Since in typical 3-dimensional (3D) metals $k_{F}$ is of
the size of the Brillouin Zone (BZ), $2k_{F}>k_{D}$, all populated
phonons can scatter off electrons. For low density electron
systems, however, the Fermi surface can be substantially smaller
than the size of the BZ, and hence $k_{F}\ll k_D$. In this case,
only a small fraction of the acoustic phonons with energies $\hbar
v_s k_{ph}\leq 2\hbar v_s k_F$ can scatter off electrons. This
phase space restriction defines a new characteristic temperature
scale for the low density e-ph scattering, the BG temperature
$\Theta_{BG}=2\hbar v_s k_F/k_B<\Theta_{D}$. It was explicitly
shown in low density 2-dimensional (2D) electron gases formed in
semiconductor heterojunctions~\cite{Stormer90} that $\rho(T)$
drops precipitously at temperatures below $T<\Theta_{BG}$ rather
than $\Theta_{D}$. However, the 3D nature of the phonons in the
host material, and the low level of tunability of $n$, make this
system ineligible for studying 2D BG physics and the explicit
density dependence of $\Theta_{BG}$.

The advance of graphene~\cite{Geim07, Geim08} brings new aspects to the study of e-ph interactions in a low dimensional system. Graphene has a Debye temperature $\Theta_D \approx 2300$K almost an order of magnitude higher than for typical metallic systems, and the electrostatic tunability of $k_{F}\propto \sqrt{n}$ allows for a wide range of control of $\Theta_{BG}$. In addition, the single atomic plane structure of graphene provides not only a strictly 2D electronic system, but a 2D acoustic phonon system as well. These unique properties have been considered theoretically, leading to the prediction that graphene exhibits a smooth crossover behavior between the high temperature $\rho(T)\sim T$ and the low temperature $\rho(T)\sim T^4$ dependence\cite{Hwang08}. The slower reduction of $\rho(T)$ at low $T$ as compared to the $T^5$ dependence observed in typical 3D conductors can be understood by the reduced spatial dimensionality.

\begin{figure}[tbp]
\includegraphics[width=1.0\linewidth]{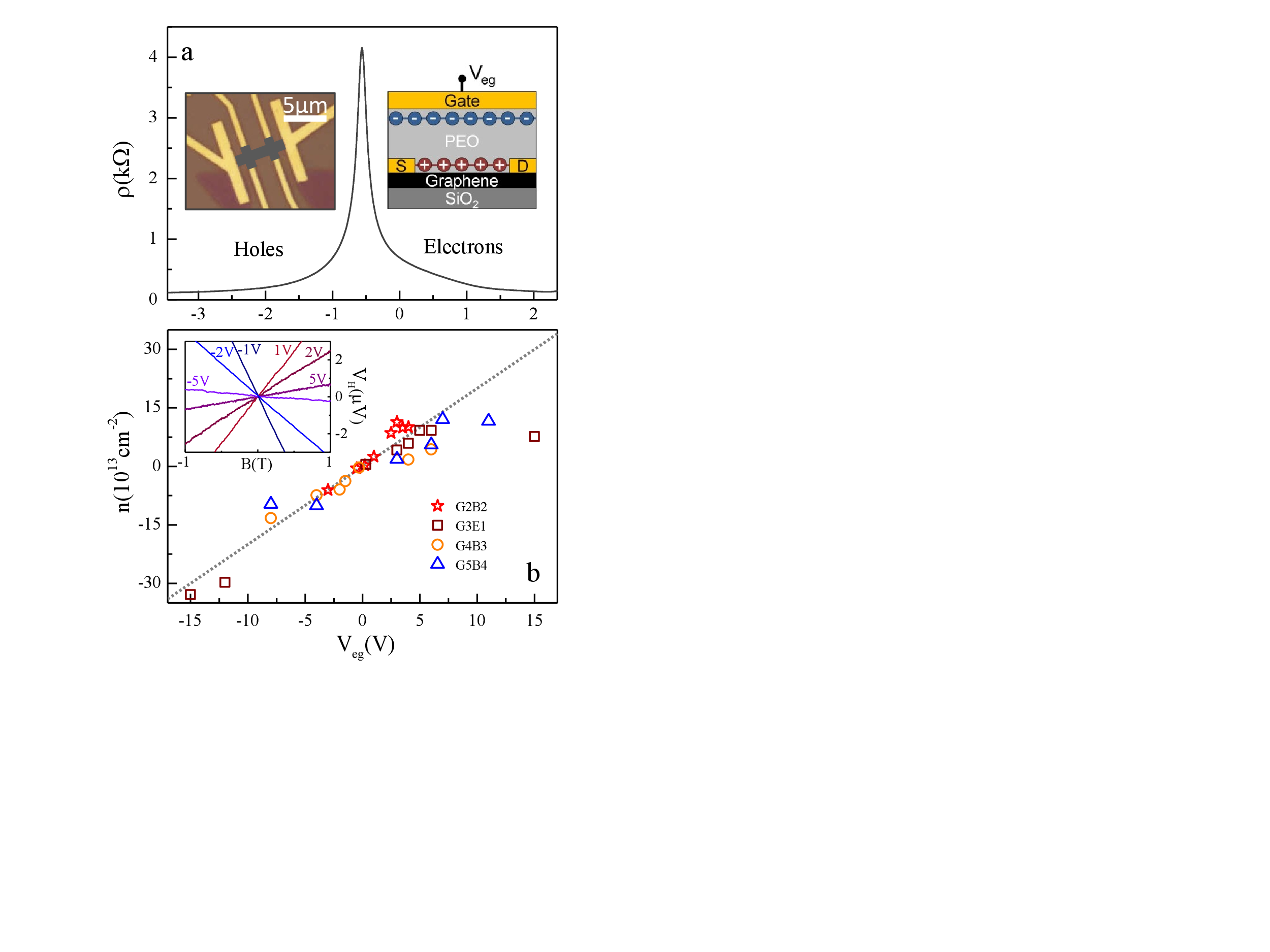}
\caption{(a) Resistivity as a function of applied electrolyte gate
voltage $V_{eg}$ at $T=300$~K. Right inset: a schematic view of the
electrolyte gated device. The Debye Layers are formed $d\sim$
1~nm above the graphene surface. Left inset shows an optical
microscope image (false colors) of a typical etched Hall bar device. (b) Inset
shows the Hall voltage $V_H$ as a function of magnetic field $B$ for different $V_{eg}$ (current across sample $I=$100~nA). The
main panel shows the extracted densities $n$ from the various Hall measurements
as a function of $V_{eg}$ for 4 different samples. The slope of the line fit represents the capacitive coupling of the electrolyte gate to the graphene.} \label{Fig.1}
\end{figure}

In this letter, we report the experimental observation of a 2D
BG behavior in graphene. Using an electrolytic gate, we
achieve extremely high carrier densities up to
4$\times$10$^{14}$cm$^{-2}$ for both electrons and holes, tuning $\Theta_{BG}$
to values of up to $\sim$1000~K. In the low $T$ limit, $T\ll \Theta_{BG}$, a $\rho(T)\sim T^4$ is observed, reflecting the 2D nature of the electrons and the acoustic phonons in graphene. At high temperatures, the resistivity shows a semiclassical $\rho(T)\sim T$ behavior. From analysis of the experimental data of $\rho(T)$, we obtain $\Theta_{BG}(n)$ and show that $\rho(T)$ scales as a universal function $\rho(T/\Theta_{BG})$ of the normalized temperature $T/\Theta_{BG}$ for all densities $|n|$, which agrees well with the theoretical expressions.

In previous graphene experiments employing thermally grown SiO$_2$
layers as the gate dielectric~\cite{Chen08,Morozov08}, $\rho(T)$
could be measured only in the density range $|n|<5 \times
10^{12}$~cm$^{-2}$. In this range of carrier densities,
$\rho(T)\sim T$ was reported for all $T<~$150~K, and, at higher
temperatures, $\rho(T)$ exhibited a rapid increase presumably due
to the scattering by thermally activated SiO$_2$ polar optical
phonons~\cite{Chen08}, by thermally quenched graphene
ripples~\cite{Morozov08}, or by Coulomb impurities~\cite{Hwang09}.
These extrinsic effects become less pronounced at higher carrier
densities where the carrier screening is enhanced~\cite{Chen08}.
Furthermore, an increased carrier density would result in an
increase of $\Theta_{BG}$, allowing to access the non-linear
$\rho(T)$ of the BG regime in a much wider temperature range.

In order to increase the electrostatic doping, we employ a solid polymer electrolyte gate~\cite{Frisbie07,Iwasa08}. This approach has previously been applied to graphene samples~\cite{Das08,
Fai09, Pinczuk09} and carrier densities of
$n\sim$~10$^{13}$~cm$^{-2}$ have been reported under ambient
conditions. Following this experimental approach, we increase the efficiency of the electrolyte gate by employing a rapid cooling method which prevents sample degradation, and reach $n>10^{14}$~cm$^{-2}$ for both electron and holes.

Fig. 1(a) shows a working principle of the solid polymer
electrolyte gate used in our experiment. Li$^{+}$ and
ClO$_{4}^{-}$ ions are mobile within a solid "mesh" formed by the
polymer poly(ethylene)oxide (PEO). By applying a voltage $V_{eg}$
to the electrolyte gate, the ions form Debye layers on top of the graphene and the
gate electrode, respectively. The extreme proximity of these charged layers, separated only by the Debye length $\lambda_{D}\sim$ 1~nm from the graphene surface, results in huge capacitances per unit area
$C_{eg}=1/\epsilon\lambda_D$. Under ambient conditions the electrolyte gate $V_{eg}$ can
be swept continuously, doping the graphene samples to either
electrons or holes, inducing a modulation of $\rho(V_{eg})$ Fig.~1(b).

The maximal $|n|$ that can be induced by the solid polymer electrolyte is mainly limited by the on-set of electrochemical reactions of the ions with the graphene, which typically turn on when $V_{eg}\gtrsim3$~V are applied. Although the threshold of the electrochemistry,
signaled by a steady increase of $\rho$ with time, depends on the
details of the particular device and sample quality, the time span until complete degradation of the sample is typically just a few minutes. We could apply extremely high $V_{eg}$ of up to 15~V to the electrolyte, avoiding electrochemically induced sample degradation by immediate cooling of the sample ($<1$~min) below $T<250$~K. At this temperature, both, the Li$^{+}$
and ClO$_{4}^{-}$ ions "freeze" out and are no longer mobile within the PEO, fixing the accumulated charges in the Debye layers to the graphene surface. The induced charge carrier densities do not vary significantly over time and temperature until after the sample is warmed up again. Accumulated $|n|$ for each applied gate voltage
$V_{eg}$ are characterized directly by performing Hall
measurements (Fig.~1(b) inset). Fig.1~(b) shows the measured $n$ as
a function of $V_{eg}$. The capacitive coupling of the
electrolyte can be then estimated from the slope of $n(V_{eg})$
where we obtain $C_{eg} \approx~$3.2~$\mu$Fcm$^{-2}$, which is
more than 250 times higher than that for typical 300~nm thick
SiO$_{2}$/Si back gates.

In this experiment we have measured $\rho(T)$ for more than 10 single layer graphene samples,
in the temperature range $1.5 < T < 300$~K and for $|n|<2 \times
10^{14}$~cm$^{-2}$. Fig.2~(a) shows a representative data set for
the measured $\rho(T)$ at various fixed $n$. Generally, $\rho(T)$ decreases
monotonically as $T$ decreases, saturating to $\rho_0$ in the low
temperature limit~\cite{footnote1}. This residual resistance $\rho_0$
stems from almost temperature independent scattering
mechanisms, such as static impurity and point defect scattering, as discussed
in previous studies~\cite{Morozov08,Chen08}. As shown in
Fig.~2(b) inset, the corresponding mobility follows $\mu_0^{-1}=(\rho_{0} en)\approx a+bn$~\cite{DasSarma10} with the fitting parameters $a=3.3\times
10^{-4}$Vs/cm$^{2}$ and $b=3.9\times 10^{-18}$Vs, representing
long and short range impurity scattering, respectively.

\begin{figure}[tbp]
\includegraphics[width=1.0\linewidth]{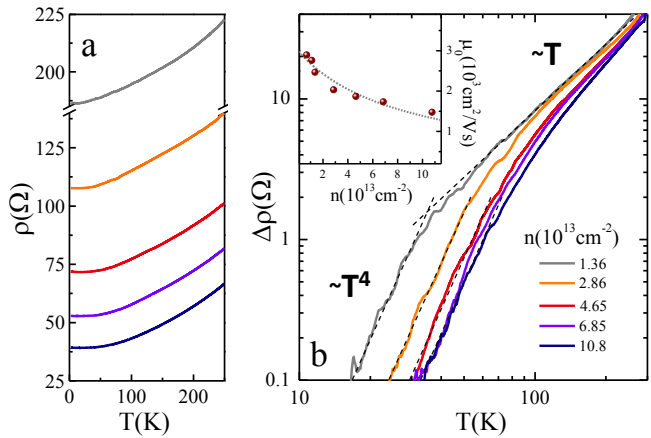}
\caption{(a) Temperature dependence of the resistivity for
different charge carrier densities of sample G8A4.(b) The temperature dependent
part of the resistivity $\Delta \rho(T)$ scales as $T^{4}$ in
the low $T$ range and smoothly crosses-over into a linear $T$
dependence at higher $T$. Dashed lines represent fits to the linear $T$ and
$T^{4}$ dependency, respectively. Inset shows the mobility $\mu_0$
at $T=2$~K as a function of the density $n$. Grey line is the theoretically expected mobility due to short and long range impurity scattering.}
\label{Fig.2}
\end{figure}

At a first glance, the temperature dependent $\rho(T)$ can be subdivided into two different temperature regimes: (i) the high temperature linear $T$ regime; and (ii) the low temperature non-linear $T$ regime~\cite{footnote2}.
This transitional trend of $\rho(T)$ at low temperatures can be better scrutinized by
subtracting off $\rho_0$ from $\rho(T)$. Fig.~2(b) displays $\Delta
\rho(T)=\rho(T)-\rho_0$ as a function of $T$ in the logarithmic
scale. At a given density $n$, each curve of $\Delta \rho$ shows a
clear transition from a linear high temperature behavior ($\rho
\sim T$) to a superlinear ($\rho \sim T^{4}$) behavior at low temperatures, as is expected from the BG model applied to electron-acoustic
phonon scattering in graphene~\cite{Hwang08}. The cross-over temperature between these two different regimes appears to be higher for higher carrier densities, in good accordance with the BG description presented above, where $\Theta_{BG}\propto\sqrt{n}$ .

We now quantitatively analyze our data in terms of the BG model. Considering the
e-ph interaction as the major source of
scattering, the temperature dependent resistivity of graphene can be obtained using the Boltzmann
transport theory~\cite{Hwang08}:
\begin{equation}
\Delta\rho (T)=\frac{8D_A^2k_F}{e^2\rho_m v_s v_F^2}f_s(\Theta_{BG}/T), \label{a1}
\end{equation}
where $D_A$ and $\rho_m$ are the acoustic deformation potential and the mass
density of graphene, respectively, $v_{F}$ the Fermi velocity and the generalized BG function for graphene
is given by the integration form:
$f_s(z)=\int_0^1\frac{zx^4\sqrt{1-x^{2}}e^{zx}}{(e^{zx}-1)^2}dx$. We remark that
Eq.~\ref{a1} is different from a typical BG formula for a 3D metal in three points.
First, the integrand contains $x^{4}$ instead of $x^{5}$, reflecting the 2D nature of
electrons and acoustic phonons in graphene. Second, the relevant
normalized temperature scale is $\Theta_{BG}$ instead of
$\Theta_D$, considering the fact $\Theta_{BG}<\Theta_D$ in our
experimental range. Third, the absence of backscattering for the carriers manifests itself in the factor $\sqrt{1-x^{2}}$ in the integrand, representing the chiral nature of the carriers in graphene~\cite{private}.

\begin{figure}[tbp]
\includegraphics[width=1.0\linewidth]{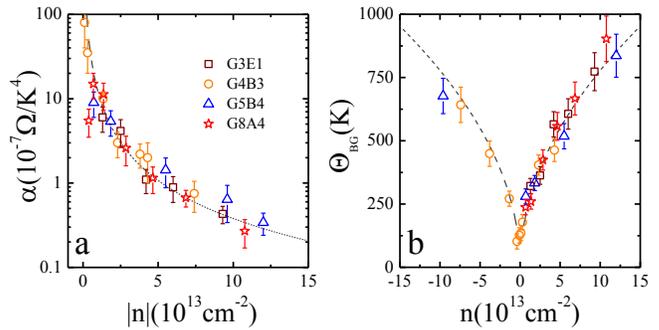}
\caption{ (a) Scaling of the prefactors $\alpha(n)$ ($\Delta\rho
\approx \alpha(n)T^{4}$). Data points were obtained from $T^{4}$
fits of the $\protect\rho$(T) traces at different carrier
densities and for different samples. The dashed line represents a
theoretically predicted fit $\propto |n|^{-\frac{3}{2}}$. (b)
$\Theta_{BG}$ at different carrier densities (symbols are defined as in (a)). The grey line is a fit to the theoretically predicted $\Theta_{BG}=2\hbar v_s \sqrt{\pi n}/k_B$.}
\label{Fig.3}
\end{figure}

Taking the two opposite limits of the temperature ranges, Eq.~\ref{a1}
can be further approximated to $\Delta\rho \approx \gamma T$ for
$T \gtrsim \Theta_{BG}$ and $\Delta\rho \approx \alpha T^4$ for $T
\ll \Theta_{BG}$, where the temperature independent proportionality
coefficients are explicitly given by~\cite{Hwang08}:
\begin{equation}
\gamma =\frac{\pi D_A^2 k_B}{4 e^2\hbar\rho_m v_s^2v_F^2} \label{a2}
\end{equation}
and
\begin{equation}
\alpha=\frac{12\zeta(4)D_A^2 k_B^4}{e^2\hbar^4 \rho_m
v_s^5 v_F^2} (\pi n)^{-\frac{3}{2}} \label{a3}
\end{equation}
where $\zeta$ is the Riemann-Zeta function.

Here we particularly note that $\alpha \propto |n|^{-3/2}$, while, $\gamma$ is
density-independent. Using these properties, we obtain $\gamma$ and $\alpha(n)$ from the experimentally observed $\Delta \rho$ at fixed $n$. First, $\gamma \approx(0.14\pm 0.01)$~$\Omega/$K is estimated from the converging high temperature limit
(dotted line in Fig.~2(b) for example). This value is in reasonable agreement with the previous
studies~\cite{Morozov08,Chen08}. We then estimate $\alpha$ from each $\rho(T)$ curve at different
densities by fitting to $\Delta\rho \sim T^4$. Fig.~3(a) shows the resulting $\alpha$ versus $|n|$ in a
wide range of experimentally accessible $|n|$ for 4 different samples. A clear trend of
$\alpha(n)\sim |n|^{-3/2}$ can be seen (dashed trace in accordance with Eq.~\ref{a3}).

The combination of the two coefficients $\alpha(n)$ and $\gamma$ allows us to compute the ratios of
$D_{A}^{2}/v_s^{2}$ and $D_{A}^{2}/v_s^{5}$, respectively, and
thus evaluate the values of $D_{A}$ and $v_{s}$ separately.
Employing $\rho_{m}=7.6\times 10^{-7}$~kg/m$^{2}$, and
$v_{F}=10^6$~m/sec, we find that the average values for each
parameter are $v_s=(2.6\pm0.4)\times10^4$~m/sec and $D_{A}=(25\pm5)$~eV, in a good agreement with the values from the literature~\cite{Hwang08,Chen08,vs,D}. Using these we can now fit each experimental curve $\Delta \rho(T)$ by Eq.~1, using $\Theta_{BG}$ as a single fitting parameter. Fig.~3(b) displays the
experimentally determined $\Theta_{BG}$ from this fits as a function of $n$ for
both electrons and holes for all measured samples. The obtained $\Theta_{BG}$ exhibit
the predicted $\sqrt{|n|}$ dependence, explicitly demonstrating our capability to tune the
BG temperature of up to $\sim$1200~K with the solid polymer electrolyte gate.

We finally discuss the universal scaling of $\Delta\rho(T)$.
The experimental estimation of $\Theta_{BG}$ allows us now a direct
test of the scaling behavior of $\Delta\rho(T)$ following Eq.~\ref{a1} in all temperature ranges. Taking
$T/\Theta_{BG}$ as a dimensionless parameter, Eq.~\ref{a1} can be
rewritten in a dimensionless scaling form:

\begin{equation}
\frac{\Delta\rho (T)}{\Delta\rho (\xi\Theta_{BG})}=
\frac{f_s(\Theta_{BG}/T)}{f_s(\xi^{-1})} \label{a4}
\end{equation}

where $\xi$ is an arbitrary constant setting the normalization temperature relative to
$\Theta_{BG}$. Fig.~4 displays the universal scaling behavior of
Eq.~\ref{a4} converted from Fig.~2. Here we simply choose $\xi=0.2$ to
ensure $\Delta\rho(\xi\Theta_{BG})$ is within the experimentally
accessible range. Remarkably, each normalized curve of $\Delta
\rho$ with different $n$ (thus different $\Theta_{BG}$) falls on top of the theoretical curve, indicating the BG model of
e-ph scattering fully explains $\rho(T)$ not only in the low
and high temperature limits, but for all temperatures.

\begin{figure}[tbp]
\includegraphics[width=1.0\linewidth]{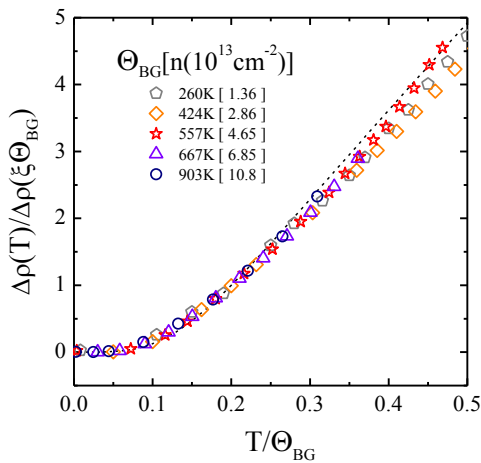}
\caption{Universal scaling of the normalized resistivity $\Delta \rho (T)/\Delta \rho (\xi\Theta_{BG})$ as a function of the normalized temperature $T/\Theta_{BG}$, explicitly using the constant $\xi=0.2$. Data points correspond to $\Delta\rho(T)$ of sample G8A4 at different $n$ and $\Theta_{BG}$ and are normalized with respect to $\Delta \rho (\xi\Theta_{BG})$ --- the resistivity at $T=\xi\Theta_{BG}$. The dashed line trace represents the theoretically predicted scaling of $f_s(\Theta_{BG}/T)/f_s(\xi^{-1})$ without use of fitting parameters.}
\label{Fig.4}
\end{figure}

In conclusion, using the electrolyte we have achieved extremely high
carrier densities of up to $|n|=4 \times 10^{14}$~cm$^{-2}$ in graphene samples. This advancement allowed us to observe a strictly 2D Bloch-Gr\"{u}neisen behavior in the measured resistivity, exhibiting the linear $T$ to superlinear $T^{4}$ cross-over, defined by the gate tunable characteristic temperature $\Theta_{BG}$. Our quantitative analysis of the temperature dependent resistivity shows an universal scaling behavior of the normalized resistivity $\rho(T)$ with the normalized temperature $T/\Theta_{BG}$, representing the 2D nature of the electrons and phonons along with the chiral nature of the carriers in graphene.

The authors thank E. H. Hwang, I.L. Aleiner, S. Das Sarma and K.B. Efetov for helpful
discussions, K.F. Mak and S. Glinskis for sample preparation. This work is supported by the AFOSR MURI, FENA, and DARPA
CERA. Sample preparation was supported by the DOE (DE-FG02-05ER46215).


\begin{thebibliography}{99}


\bibitem{Abrikosov} A.A. Abrikosov, \textit{Fundamentals of the Theory of
Metals,} North-Holland (1988).

\bibitem{Bloch} F. Bloch, Z. Phys. \textbf{59} 208 (1930).

\bibitem{Grueneisen} E. Gr\"{u}neisen Ann. Phys., Lpz. \textbf{16} 530 (1933).

\bibitem{Meisner1935} W. Meissner, Handb. D. Exp. Phys. \textbf{11} 338 (1935).


\bibitem{Stormer90} H. L. Stormer, L. N. Pfeiffer, K. W. Baldwin, and K. W.
West, Phys. Rev. B \textbf{41}, 1278--1281 (1990).


\bibitem{Geim07} A.~K. Geim and K.~S. Novoselov, Nat Mater \textbf{6}, 183
(2007).

\bibitem{Geim08} A.~K. Geim and P. Kim, Scientific American \textbf{298}, 68
(2008).

\bibitem{Hwang08} E.~H. Hwang and S. Das Sarma, Phys. Rev. B \textbf{77},
115449 (2008).

\bibitem{Morozov08} S. V. Morozov, K. S. Novoselov, M. I. Katsnelson, F.
Schedin, D. C. Elias, J. A. Jaszczak, and A. K. Geim, Phys. Rev. Lett.
\textbf{100}, 016602 (2008).

\bibitem{Chen08} J.~H. Chen, C. Jang, S. Xiao, M. Ishigami, M. S. Fuhrer,
Nature Nanotechnology \textbf{3}, 206 - 209 (2008).

\bibitem{Hwang09} E. H. Hwang and S. Das Sarma, Phys. Rev. B \textbf{79}, 165404 (2009).

\bibitem{Frisbie07} Matthew J. Panzer, C. Daniel Frisbie, Advanced Materials
\textbf{20}, 3177 - 3180 (2008).

\bibitem{Iwasa08} K. Ueno, S. Nakamura, H. Shimotani, A. Ohtomo, N. Kimura,
T. Nojima, H. Aoki, Y. Iwasa and M. Kawasaki, Nature Materials \textbf{7},
855 - 858 (2008).

\bibitem{Das08} A. Das, S. Pisana, B. Chakraborty, S. Piscanec, S. K. Saha,
U. V. Waghmare, K. S. Novoselov, H. R. Krishnamurthy, A. K. Geim, A. C.
Ferrari and A. K. Sood, Nature Nanotechnology \textbf{3}, 210 - 215 (2008).

\bibitem{Fai09} Kin Fai Mak, Chun Hung Lui, Jie Shan, and Tony F. Heinz,
Phys. Rev. Lett. \textbf{102}, 256405 (2009).

\bibitem{Pinczuk09} J. Yan, T. Villarson, E. A. Henriksen, P. Kim and A.
Pinczuk, Phys. Rev. B \textbf{80}, 241417(R) (2009).


\bibitem{footnote1} In our experiment, $\rho_0$ is determined from
the measured $\rho$ at the lowest temperature ($\approx 2$~K).

\bibitem{DasSarma10} S. Das Sarma, E. H. Hwang, E. Rossi, Phys. Rev. B \textbf{81}, 161407 (2010).

\bibitem{footnote2} For $|n| \lesssim 10^{13}$~cm$^{-2}$, a strong activation behavior of $\rho(T)$ is observed for $T\geq 150$~K as shown in previous studies~\cite{Morozov08,Chen08}.

\bibitem{private} E. H. Hwang (private communication).

\bibitem{vs} L. Pietronero, S. Str\"{a}ssler, H. R. Zeller, and M. J. Rice, Phys. Rev. B \textbf{22}, 904 (1980); L. M. Woods and G. D. Mahan, Phys. Rev. B \textbf{61}, 10651–10663 (2000); H. Suzuura and T. Ando, Phys. Rev. B \textbf{65}, 235412 (2002); G. Pennington and N. Goldsman, Phys. Rev. B \textbf{68}, 045426 (2003).

\bibitem{D} S. Ono and K. Sugihara, J. Phys. Soc. Jpn. \textbf{21}, 861 (1966); K. Sugihara, Phys. Rev. B \textbf{28}, 2157 (1983).


\end{thebibliography}
\end{document}